\documentclass[prl,
twocolumn,
aps,psfig]{revtex4} %/MH

\usepackage{graphicx,color,epsfig,rotate}

\begin{document}
\draft

%\wideabs{

\title{Optical Study 
of 
LaO$_{0.9}$F$_{0.1}$FeAs: Evidence for a Weakly Coupled Superconducting State}

\author{ S.-L. Drechsler$^{*}$, 
M.~Grobosch, K.~Koepernik, 
G.~Behr, A.~K\"ohler, J.~Werner, A.~Kondrat, N.~Leps, 
C.~Hess, R.~Klingeler, R.~Schuster, B.~B\"uchner, and M.~Knupfer}

\affiliation{IFW Dresden, P.O. Box 270116, D-01171 Dresden,
Germany}

\date{\today}
\begin{abstract}
We have studied the reflectance of the recently discovered
superconductor LaO$_{0.9}$F$_{0.1}$FeAs in a wide energy range
from the far infrared to the visible regime. We report on the
observation of infrared active phonons, the plasma edge (PE) and
possible interband transitions.
 On the basis of this
data and the reported in-plane penetration 
depth $\lambda_L(0)=$~ 254~nm 
[H.\ Luetkens {\it et al.}, Phys.\ Rev.\ Lett.\ {\bf 101}, 097009 (2008)] a 
disorder sensitive relatively
small value of the
total electron-boson
coupling constant 
$\lambda_{tot}=\lambda_{e-ph}+\lambda_{e-sp}\sim$ 
~0.6 $\pm~0.35$ 
can be estimated
adopting 
an effective single-band 
%clean limit 
picture.

\end{abstract}

%\pacs{78.20.-e,71.35.-y,78.90.+t}

\maketitle

The discovery of superconductivity up to 43 K 
in LaO$_{1-x}$F$_{x}$FeAs 
\cite{kamihara08,takahashi08} 
has established a new
family of superconductors with rather high transition
temperatures $T_c$. 
The substitution of La with other rare earths even
 yields 
$T_c$-values
 above 
50~K 
\cite{ren_a}. In addition 
to these already fascinating reports, both
experimental and theoretical 
studies consider possible
unconventional multiband behavior \cite{cao,eschrig}.
%and evidence for gap nodes \cite{shan,mu,ren_e}, while
Various scenarios for the superconducting mechanism 
have been proposed or excluded \cite{boeri,mazin,lee,han,dai,kuroki,xu}.
A more general related problem under 
debate \cite{cao,haule,shorikov,si,daghofer,weng,haule2,kurmaev} 
being of 
considerable interest
is the strength of the correlation effects in these doped pnictides
governed by the local Coulomb interactions 
$U_{d}$ and $J_{d}$
%-values  
at Fe sites.

The crystals of LaO$_{1-x}$F$_{x}$FeAs are
 formed by
alternating stacking of LaO$_{1-x}$F$_{x}$ and FeAs layers. 
The
 electronic structure of these new systems 
has been mainly
 addressed by theoretical studies so far. They
predict
that LaOFeAs and the related compounds harbor quasi
two-dimensional electronic bands as is suggested by the layered
crystal structure already \cite{eschrig,cao,boeri,mazin,kuroki,xu}
.
The \textcolor{black}{LDA (local density approximation)} 
Fermi surface of the doped compound
consists of four cylinder-like sheets
. Hence, 
the electronic behavior is expected to be highly
anisotropic concerning e.g.~charge transport and superconducting
properties such as the penetration depth 
$\lambda_L(T)$ and the upper critical magnetic field.
In addition, calculations have provided the expected phonon
density of states which might be relevant 
for the symmetry of the superconducting order parameter
even in the case
of a dominant magnetic coupling in spite of the predicted weak 
electron-phonon ({\it e-ph})
coupling
\cite{boeri,singh}. 
So far, optical data are available for the
far-infrared region only, where indications for the opening of a
gap in the superconducting state have been found, and infrared
active phonons can be observed \cite{chen_c}.
To the best of our knowledge a phenomenological estimate
of the electron-boson ({\it e-b}) coupling strengths 
%based on a thermodynamic analysis 
%of the superconducting properties 
is still missing and the present work provides a 
first step in that direction.

\indent
Here we report on a study of the optical properties of
LaO$_{0.9}$F$_{0.1}$FeAs via reflectance measurements. We have
determined the reflectance of powder samples in the energy range
from 0.009 up to 3~eV at $T$~=~300~K. Our data reveal two
prominent and two weaker infrared active phonon modes, and provide
evidence for the in-plane plasma energy at $\approx$ 0.4~eV as well
as the appearance of electronic interband transitions at higher
energies.
 These data are used to estimate 
the strength of the total {\it e-b}
coupling $\lambda_{tot}$
responsible for 
the "mass enhancement" 
in the plasma energy of the paired electrons  
without specifying the nature of this coupling.
We compare its value with various microscopic 
estimates in the literature
based on spin fluctuations
and 
%electron-phonon 
{\it e-ph}
contributions as well as the empirical 
unscreened plasma energy with uncorrelated
L(S)DA (local spin density approximation) predictions.

A polycrystalline sample of LaO$_{0.9}$F$_{0.1}$FeAs was prepared
as described in our 
previous work \cite{luetkens} and in Ref.~
\cite{zhu}. For the optical measurements reported here
the pellets have been polished to obtain appropriate surfaces. In 
order to characterize the superconducting 
properties, zero
field cooled (shielding signal) and field cooled (Meissner signal)
magnetic susceptibility in external fields $H$~=~10~Oe ... 50~kOe
have been measured using a SQUID magnetometer. 
The resistivity has
been measured with a standard 4-point geometry employing an
alternating DC current. A value of $T_c$~$ \approx $~26.8~K has been 
extracted from 
these data as shown in Fig.~1.

\begin{figure}[t]
\includegraphics[width=7.5cm,angle=0]{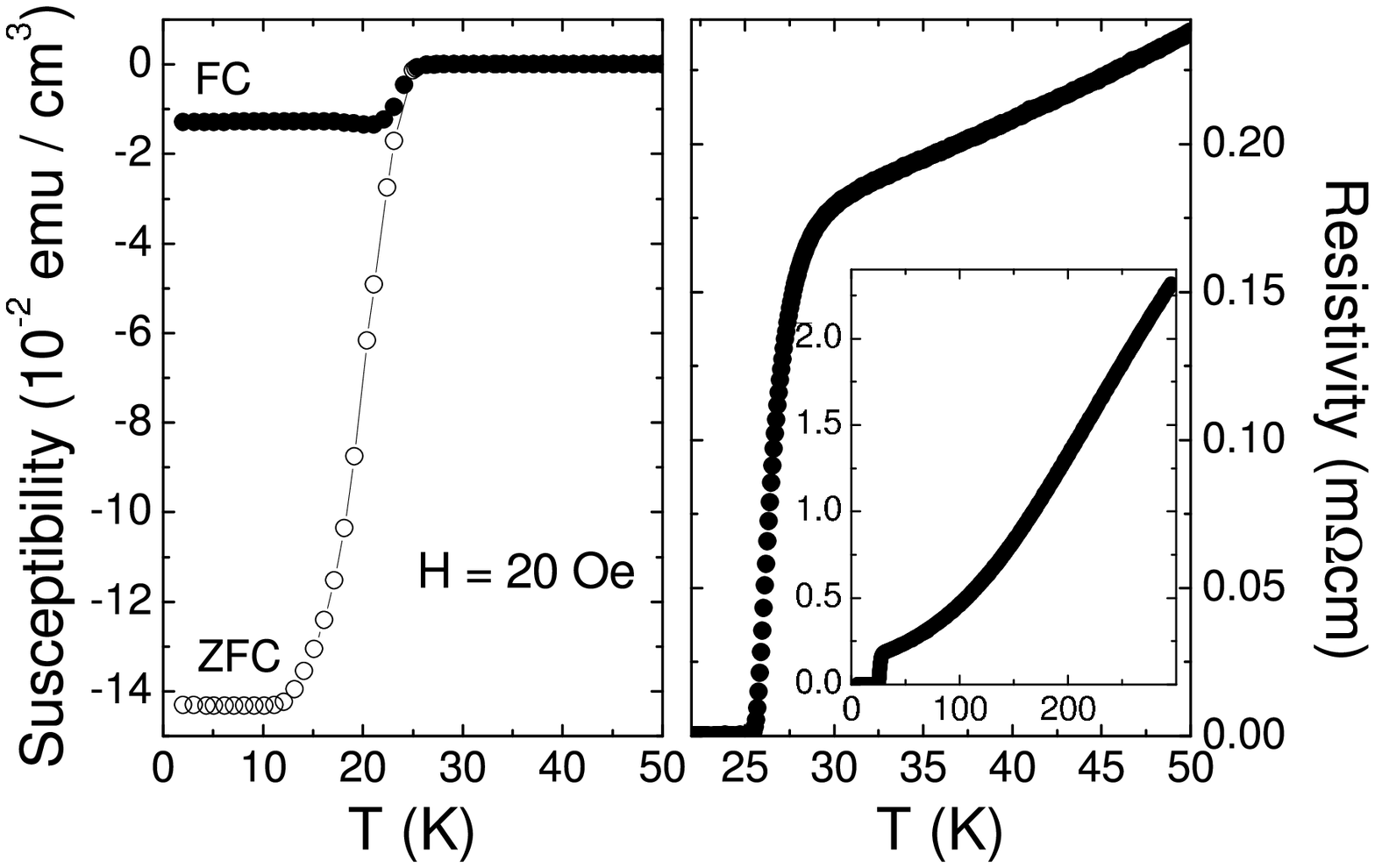}
\caption{Left: field cooled (FC) and zero field cooled (ZFC) magnetic
susceptibility for the LaO$_{0.9}$F$_{0.1}$FeAs samples 
studied here. 
Right: $T$-dependence of the resistivity near  
$T_c$.} 
\label{chirho}
\end{figure}

The reflectance measurements have been performed using a
combination of Bruker IFS113v/IFS88 spectrometers. This allows us
to determine the reflectivity from the far infrared up to the
visible spectral region in an energy range from 0.009 up to 3~eV.
The measurements have been carried out with different spectral
resolutions depending on the energy range, these vary form 0.06
 to 
2~meV. Since the observed spectral features are
significantly broader than these values, this slight variation in
resolution does not impact our data analysis. All reflectance
measurements were performed at $T=$~300~K 
and a pressure of
4~mbar.

In Fig.~2 we show the reflectance of our LaO$_{0.9}$F$_{0.1}$FeAs
sample (note the logarithmic and linear energy scales in the two
panels, respectively). These data reveal a number of spectral
features with different origin. At low energies (see left panel of
Fig.~2) there are two prominent structures at about 12 and 55~meV
and two weaker features in-between near 25 and 31~meV,
which most likely stem from the excitation of infrared active
phonons. We attribute the two strong features to
phonons that are polarized along the {\bf c}-axis of
the lattice, since the dielectric screening in this direction is
expected to be much weaker as a result of the rather two
dimensional electronic system. An assignment of the two weaker
phonons is not possible at the present stage, and future work on
single crystals will help to clarify their polarization character.
A comparison with a calculated  phonon density of
states \cite{boeri} allows an assignment of the higher energy
phonon (55~meV) to oxygen modes, since above about 40~meV only
these vibrations can be expected. In addition, the calculated
phonon density of states has a maximum near 12~meV \cite{boeri}
with contribution from all elements in the structure, which would
correspond to the strong lowest energy feature in Fig.~2, and the
calculations also predict two further maxima which could be
associated with weaker structures in our reflectance spectrum.
\begin{figure}[b]
\includegraphics[width=8cm]{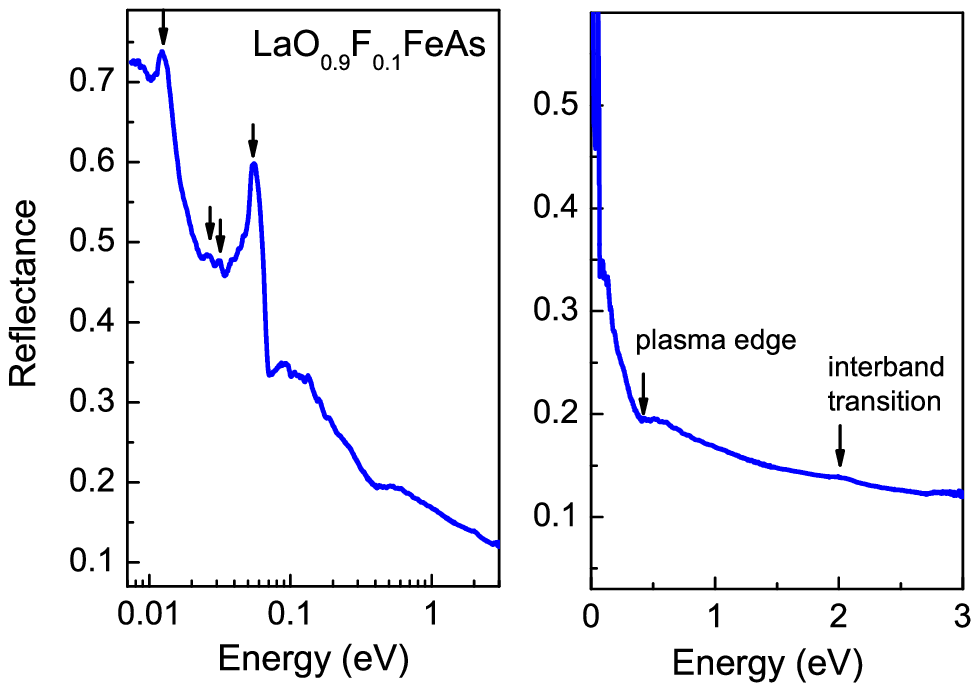}
\caption{Reflectance of a polished
LaO$_{0.9}$F$_{0.1}$FeAs sample. The energy is given in a
logarithmic (left) and linear (right) scale. The
arrows in the left highlight the low-energy excitation features 
ascribed
%that are most likely related 
to phonons. In the right the arrows denote
the plasma edge and the 2~eV interband transition.} \label{f3}
\end{figure}

At higher energies the reflectance significantly drops between 0.13
 and 0.4~eV, thereafter it shows a small increase until about
0.6~eV before it starts to slightly decrease again. In addition,
at about 2~eV another small upturn is visible. We attribute the
two small features at about 0.6 and 2~eV to weak electronic
interband transition at the corresponding energies. 
\textcolor{black}{This observation is in accord with the behavior
of the optical conductivity $\sigma(\omega)$ predicted by 
Haule {\it et al.} using dynamical mean-field theory (DMFT), 
where As 4$p$ to  Fe 3$d$ interband
 transition (IB$_{pd}$) 
slightly below 2~eV and a weak feature  near 0.6~eV
for the undoped parent compound LaOFeAs have been found.}
Again, our
measurements on powder samples do not allow the extraction of
their polarization. The steepest edge centered at about 250~meV
with an high energy onset at about \textcolor{black}{395(20)}~meV 
represents the plasma
edge or plasma energy of LaO$_{0.9}$F$_{0.1}$FeAs, which is
observed at a relatively low value. In consideration of the quasi-2D
character of the electronic states in
LaO$_{0.9}$F$_{0.1}$FeAs and the expected strong anisotropy of the
plasma energy for light polarized $\parallel$ to the ({\bf a,b})
and {\bf c} crystal axes, respectively, the onset of the plasma
edge (PE) in Fig.~2  gives a reasonable value of the
in-plane value (i.e.~for light polarized within the ({\bf a,b})
crystal plane) \cite{remarkpoly}.

Within a simple Drude model, this value allows a first, crude
estimate of the charge carrier density in
LaO$_{0.9}$F$_{0.1}$FeAs. Within this model, the plasma energy,
$\omega_p$, is given by $\omega_p^2 = ne^2/m \varepsilon_0
\varepsilon_\infty$, with $n$, $e$, $m$, and $\varepsilon_\infty$
representing the charge carrier density, the elementary charge,
the effective mass of the charge carriers and the background
dielectric screening due to higher lying electronic
excitations.\cite{wooten} Taking the effective mass equal to the
free electron mass, and $\varepsilon_\infty \sim $ 10-12 \cite{boris}, 
one would arrive
at a rather low charge density of about 
4x10$^{-20}$ cm$^{-3}$. We
note, that this estimate ignores the strongly anisotropic nature
of the electronic 
%orbitals 
structure
near the Fermi energy $E_F$ 
in LaO$_{0.9}$F$_{0.1}$FeAs.
Our LDA-FPLO [full potential localized orbital
minimum-basis code) (version 7)] \cite{koepernik} 
band structure calculations for the 
nonmagnetic ground 
state provide 
a 
%value for the
plasma frequency within the ({\bf a,b}) crystal plane of 
$\Omega^{LDA}_p
=2.1$~eV 
and a small value for a polarization along
the {\bf c} axis of 0.34~eV in accord with
2.3~eV and 0.32~eV given 
in Ref.\ \onlinecite{boeri}. Note that these "bare" values do not
describe the screening through $\varepsilon_{\infty}$
caused by interband transitions.
In order  to get the measured $\omega_p$~=~0.395~eV
an unusually large value of
$\varepsilon_{\infty} \sim $~28.3 to 33.9 would be required 
which seems 
to be unrealistic \cite{boris}
 and instead $\varepsilon_{\infty} \sim $ 
12
%4 to 6 like in cuprates  
would be 
expected. Then, for the empirical unscreened plasma energy $\Omega_p$
a value about
 1.37~eV is expected on the basis
of our reflectance data. 
 For a deeper insight we calculated 
at first
the interband contribution to the dielectric function
 from
the in-plane DMFT optical conductivity between 1 and 6~eV 
for the undoped LaOFeAs 
given in Fig.~5 of Ref.~\cite{haule}. 
From its static value we obtained 
$\varepsilon_{\infty}$~=~5.4, only.
Note that generally smaller 
 
values for the onsite Coulomb repulsion $U_{d}$
would result in  
larger 
%(smaller) 
$\varepsilon_{\infty }$ values due to an energy shift of the
interband 
transition from the lower Hubbard band
and the less suppressed combined density of states
for the IB$_{pd}$ transitions which vanishes
at the metal-insulator transition expected at 
a critical
$U_d$ slightly above 4~eV \cite{haule}. 
Hence, a corresponding systematic DMFT study 
with lower $U_{d}$ and $J_{d}$-values would 
be of interest. In 
contrast to Haule {\it et al.} \cite{haule},  
Shorikov {\it et al.} \cite{shorikov} argue that LaOFeAs is in an 
intermediate $U_{d}$ regime  
but strongly affected by the value of the 
Hund's rule exchange $J_{\mbox{\tiny Fe}}$.
Hence, 
$\varepsilon_{\infty}$ should significantly
increase and the $U_d$= 4~eV DMFT based 
estimate given above can be regarded as
a {\it lower} 
limit. But it provides an unrealistic estimate for 
 $\varepsilon_{\infty}$
 itself. The expected  $U_d$-dependence of 
$\varepsilon_{\infty}$ is shown schematically in Fig.\ 3. It points to
strongly reduced $U_d$-values in accord with the suggestions
of Refs.\ \cite{shorikov,kurmaev,sawatzky}. Thus, $\varepsilon_{\infty}$
provides a convenient direct measure for the strength of the 
correlation regime, even for small $U_d$-values near 2 eV.

In addition, there is a direct relation between the plasma
energy as measured with optical techniques and the London
penetration depth in the superconducting state. Within a BCS approach, 
these two parameters are inversely proportional
to each other. Recently, the in-plane penetration depth,
$\lambda_{a,b}$, for LaO$_{0.9}$F$_{0.1}$FeAs has been determined
using muon spin relaxation to be 
$\lambda^{a,b}_L$(0)~=~254(2)~nm \cite{luetkens,remark2}. This value  
slightly 
exceeds those found for optimally doped 
high-$T_c$ cuprates (HTC's). 
This suggests
that the plasma energy of LaO$_{0.9}$F$_{0.1}$FeAs should be
smaller than that of HTC. 
Their $\omega_p$
is found near 1 eV \cite{nucker89,uchida91,zibold93,knupfer94}
for optimal doping (i.e.~the highest $T_c$),
and taking into account the different penetration depths
$\lambda^{a,b}_L$ one would expect a plasma energy in
LaO$_{0.9}$F$_{0.1}$FeAs that is close to the measured value.
Here we take into account that $\lambda_L$
is renormalized by the mass enhancement factor as a
consequence of the 
always present $e$-$b$
coupling measured by the Fermi surface averaged coupling 
constant $\lambda_{tot}$, whereas the plasma energy is
not. This asymmetry in the renormalization
is based on very different temperatures and energies probed in both
measurement: nearly $T=0$ and low-$\omega$
%frequencies 
for the penetration
depth {\it vs.} high energies and high $T$ for the PE, see also Fig.~7
in Ref.~\onlinecite{basov} and
the {\it Note added}. In other words, we employ both the $T=0$ and the 
asymptotic high-energy limits
of the coupling constant being in general $\omega$- and $T$-dependent. 
In the effective single-band approach this relation 
can be rewritten in convenient units as   
 \cite{walte}
\begin{equation}
\alpha \Omega_p \mbox{[eV]}=\sqrt{(n/n_s(0))(1+\lambda_{tot}(0))(1+\delta)},
\end{equation}
where $\alpha=\lambda_L(0)$[nm]/(197.3~nm) is the  experimental 
$(a,b)$-plane penetration 
depth extrapolated to $T=0$,
$\Omega_p=\sqrt{\varepsilon_{\infty}} \omega_p $ denotes 
the empirical unscreened plasma 
energy, $n_s(0)$ is the density
 of
electrons in the condensate at $T=0$ 
and $n$ the total electron density
which contributes to the unscreened $\Omega_p$,
 $\delta=0.7\gamma_{imp}/(2\Delta(0))$ is the disorder parameter which
vanishes in the clean limit. 
Notice that all three factors under the root are $\geq$~1.
In the quasi-clean limit $\delta < 1$
at $T=0$, for $n$=$n_s$  and using 
$\varepsilon_{\infty}$~= 12 \cite{boris}, $\delta=0.93$ K for 
$\gamma_{imp}=125$ K estimated from resistivity data \cite{kulic} 
and $\lambda_L(0)$~=~254~nm 
\cite{luetkens,remark2} one estimates 
$\lambda_{tot} \sim 0.61$, i.e.
\ the
superconductivity is in a weak coupling regime 
for our samples with 
$T_c$~=~27~K.
The effect of 
$\varepsilon_{\infty}/(1+\delta )$ on 
the coupling strength 
is shown in Fig.~3.
\begin{figure}[b]
\begin{minipage}{0.98\textwidth}
\vspace{-0.5cm}
\hspace{-9cm}
\begin{minipage}{0.28\textwidth}%
\includegraphics[width=\textwidth]{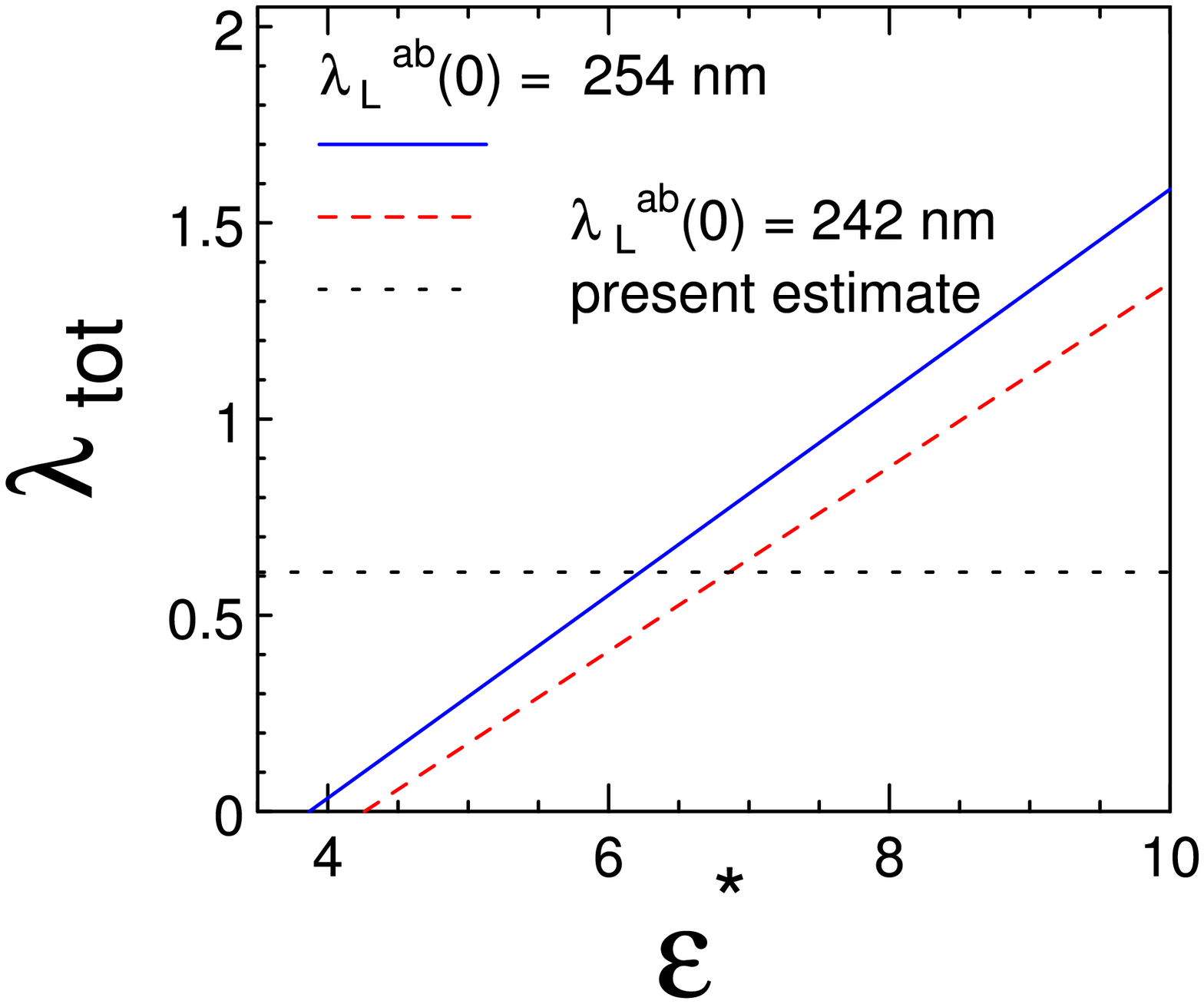}%\par
\end{minipage}
\begin{minipage}{0.20\textwidth}%
\begin{minipage}{\textwidth}%
\includegraphics[width=\textwidth]{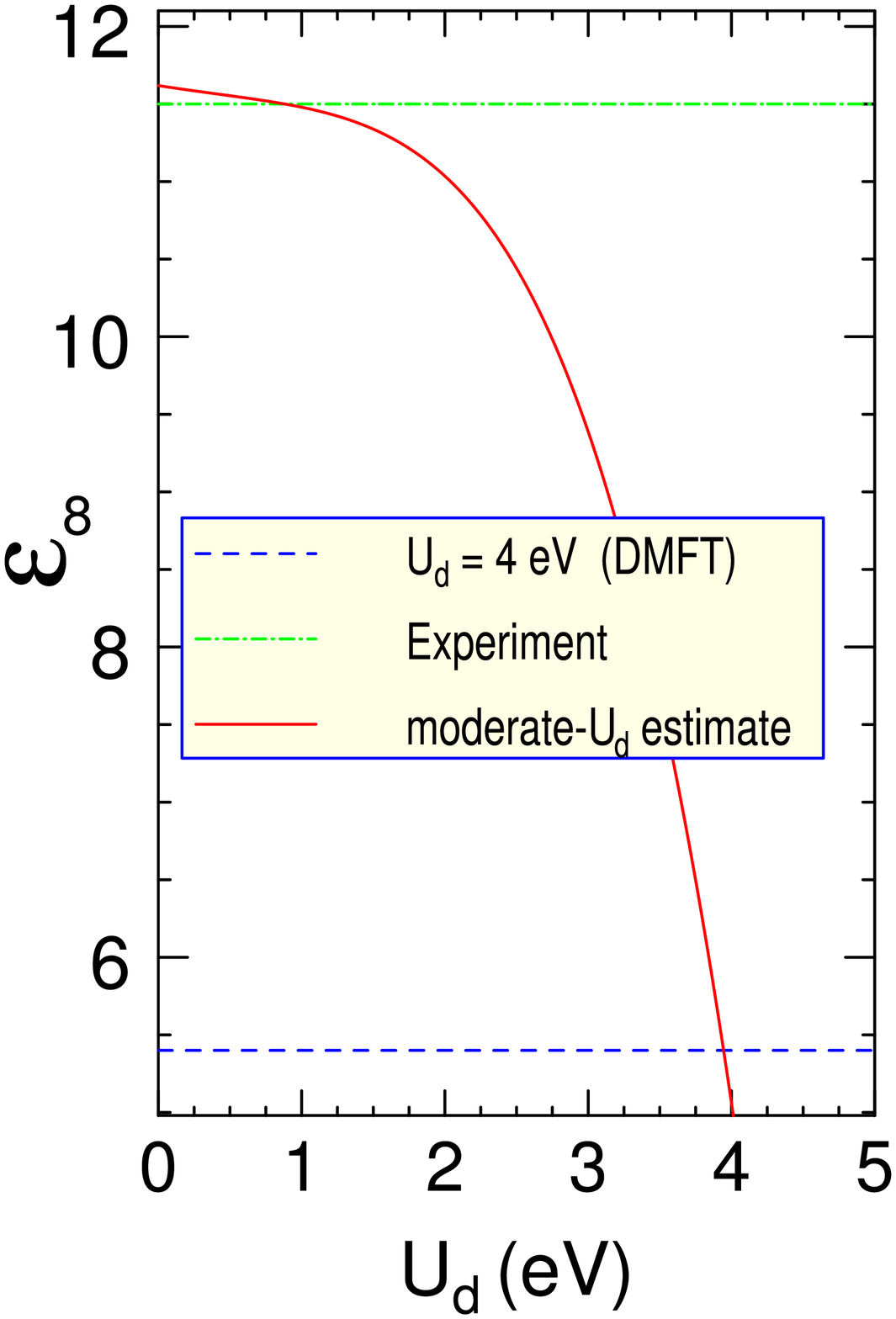}%\par
\end{minipage}
\end{minipage}
\end{minipage}
\caption{Empirical relation between 
the total coupling constant from the mass enhancement
entering the penetration depth  vs interband screening 
modeled by the
 disorder renormalized
dielectric background constant 
$\varepsilon^*=
\varepsilon_{\infty}n_s/n(1+\delta )$.
The $\lambda^{ab}_L(0)$ values from 
Refs.\ \cite{luetkens,remark2} (left).
The improved screening measured by the background dielectric
constant $\varepsilon_{\infty}$ vs.\ reduced onsite repulsion 
$U_d$ on 
Fe-sites (right).
} 
\label{schueps1}
\end{figure}
Note that a substantial
impurity scattering  ($\delta >1)$ beyond our 
moderate disorder
regime should
further 
reduce $\lambda_{tot}$. 
We note once more that our empirical $\Omega_p \approx 1.37$~eV 
 differs clearly 
from the LDA-prediction of 2.1~eV for the 
nonmagnetic ground state and slightly from
1.6~eV for the rigid band
estimate of the 
antiferromagnetic state within the LSDA. 

The microscopic nature of the empirically estimated 
total $e$-$b$ coupling constant and especially
its decomposition into the interaction with various boson modes  
remains unclear.
Since the experimental optical mass
deviates 
by a factor of 2.35
from the LDA predictions 
based on the neglect of antiferromagnetic 
fluctuations and correlation effects 
(and the LSDA by a factor of 1.37)
it is also unclear to what extent the 
calculated weak $e$-$ph$ coupling $\lambda^{LDA}_{e-ph} \leq$~0.21
\cite{boeri} can be trusted. These 
deviations might be
 caused by the filling
of the two(three) Fermi surface hole pockets \cite{haule} 
as a result of correlations
,
a hidden spin density wave \cite{weng}, or another 
still unknown many-body effect. 
Adopting nevertheless such a weak value for
$\lambda_{e-ph}$
one is left with a non-phonon contribution of 0.2 to 0.4.
Since the screening by the interband transitions is affected
by the value of $U_d$ as mentioned above,  
Fig.~3 suggests the stronger the correlation the weaker is the
empirical
{\it e-b} coupling strength.

In general, a  classical $e$-$ph$-mechanism seems to be 
ruled out by
the weakness of $\lambda_{ph}$ and the smallnes of typical phonon 
energies 
 except an exotic situation with an attractive 
Coulomb interaction $\mu^*<0$(!) 
due to a {\it negative} dielectric response at large wave 
vectors \cite{tesanovich} or huge pnictide polarizibilities
\cite{sawatzky}.   
Multi-band effects ignored here might lead to 
somewhat larger $\lambda$-values
since the low-$T$ penetration depth and 
$\lambda_L(0)$ in particular, might be affected 
by the weakest
coupled group of electrons with the smallest gaps and the largest Fermi 
velocities ($\Omega_p$) whereas the 
region near  $T_c$ 
is dominated by particles with the largest gaps \cite{golubov}. 
The corresponding estimates e.g.\ within a two-band model are 
postponed to the time when more data 
%of the underlying 
%electronic
%structure, various gaps, coupling constants, etc.\ 
will be available.

To summarize, we have studied the optical response of 
LaO$_{0.9}$F$_{0.1}$FeAs powder samples in a wide energy range. Our
reflectance
data reveal four features near 12, 25, 31, and 
55~meV we assigned to phonon excitations.
% in this material. 
Furthermore, our data allow to extract the screened in-plane
plasma energy to about 0.4~eV and two interband excitations
near 0.6 and 2~eV. The subsequent 
analysis of the reported penetration depth suggests a quasi-clean limit
,
weak 
coupling 
superconducting
regime. 
The analysis of $\varepsilon_{\infty}$ 
in the correlated regime at
$U_d$=4~eV
points to a substantially  reduced Coulomb repulsion
This is in 
accord with Refs.\ \cite{shorikov,kurmaev,sawatzky}. Further systematic 
studies 
including various dopings, oriented films, single crystals
 and other Fe based superconductors are desirable.
\\
{\it Note added.}
Preparing an amended manuscript
 we learned about an
ellipsometry study of LaO$_{0.9}$F$_{0.1}$FeAs
, 
 where $\tilde{\Omega}_p$~=~0.61~eV has 
been reported
applying the effective medium approximation (EMA) to obtain
$\sigma_{\mbox{\tiny eff}}(\omega)$ for
a polycrystalline sample \cite{boris}. However, for spherical grains and 
$\sigma_{ab}\gg \sigma_c$ as expected for a quasi-2D metal 
(except for the unscreened $c$-axis polarized i.r.~phonon related
peaks), EMA predicts 
$\sigma_{\mbox{\tiny eff}}\approx 0.5\sigma_{ab}$
which correponds to $\tilde{\Omega}^{ab}_p$~=~0.86~eV.  
Since $\tilde{\Omega}_p$ has been derived mainly from the low-$\omega$, 
damped
Drude region (below 25~meV) it is still renormalized by the $e$-$b$ 
coupling, where
a high-energy boson well above 25~meV has been assumed
to explain the high $T_c$ at weak coupling. 
Using our
$\lambda=0.61$ one estimates 
$\Omega_p=$1.23~eV, 
close to 1.37
eV suggested above \cite{remarkopt}. 
Finally, the analysis of 
the Pauli limiting (PL)
behavior for the closely related LaO$_{0.9}$F$_{0.1}$FeAs$_{1-\delta}$ 
system yields a similar value
$\lambda \approx 0.6$ to 0.7 as derived from the strong 
coupling correction
for its 
PL field $B_P$(0)~=~102~T \cite{fuchs}. Rather similar phonon-peaks as reported
here,
have been detected recently also for SmO$_{1-x}$F$_x$FeAs \cite{mirzaei}. 

We thank M.~Deutschmann, R.~M\"uller, S.~Pichl, R.~Sch\"onfelder,
R.~H\"ubel, S. M\"uller-Litvanyi  
and S.~Leger for technical support
and H.\ Rosner, 
A.\ Boris, 
M.\ Ku-li\v{c}, O.\ Dolgov, H.~Klauss, H.~Eschrig, 
A.~Koitzsch,  
V.~Gvozdikov, and 
I.~Eremin for discussions.


\begin{thebibliography}{00}
\bibitem{remarkopt} The tails of the low-$\omega$ interband transitions take 
artificially oscillator strength
of the Drude peak. This way $\Omega_p$ is somewhat underestimated.  In 
addition, 
for a strongly correlated system
as suggested there 
%in Ref.~ \cite{haule} 
a
fit with an $\omega$- and $T$-dependent $\gamma$ should be performed, 
see e.g.\ Ref.~\cite{basov}.
\end{thebibliography}
\end{document}